\documentclass[%
preprint,
 amsmath,amssymb,
 aps,
]{revtex4-2}

\usepackage{graphicx}
\usepackage{dcolumn}
\usepackage{bm}
\usepackage{hyperref}

\begin{document}
\title{Controlled locomotion of a minimal soft structure by stick-slip nonlinearity}

\author{T.\ Barois$^{*1}$, A.\ Boucherie$^{1}$, L. Tadrist$^{2}$ and H.\ Kellay$^{1}$\\
\normalsize{$^{1}$Univ. Bordeaux, CNRS, LOMA, UMR 5798, F-33400 Talence, France,}\\ 
\normalsize{$^2$Aix Marseille Univ, CNRS, ISM, Marseille, France}\\}

\begin{abstract}
We present a locomotion mechanism that uses the stick-slip transition of a soft passive structure with an internal mechanical resonance. The structure is harmonically driven by a global vertical shaking and, because of its resonance dephasing and the threshold response of stick-slip transition, it can either move forward or backward. We establish a relation for the motion acceleration threshold that we experimentally validate. We identify a non-trivial regime close to the resonance with a velocity inversion for a constant excitation frequency and an increasing driving amplitude. We finally show that we can achieve a controlled multi-modal motion by combining multiple internal resonances. 
\end{abstract}

\maketitle

The drift of an object placed on a vibrated table is a strategy that was used with a large variety of granular particles with head-tail properties in the bouncing regime. This is the case for bolt-like particles \cite{yamada2003coherent}, rods with shifted center of mass \cite{kudrolli2008swarming}, asymmetric chains \cite{kudrolli2010concentration}, dimers with asymmetric internal dynamics \cite{xu2017self}  and asymmetric plant spikes \cite{bai2013vibration}. Self-propulsion using asymmetric elements also exists with other power sources such as the Leidenfrost levitation and motion of uneven solids \cite{linke2006self,dupeux2013self}, the horizontal motion of asymmetric objects in vertical diffusion gradients \cite{allshouse2010propulsion} or Janus-like particles \cite{casagrande1989janus,walther2008janus}.  
Rectified motion even exists with particles without head-tail asymmetry such as even dimers \cite{dorbolo2005dynamics}, trimers \cite{dorbolo2009bouncing}, or chiral dimers \cite{kubo2015mode}. Rectified motion under vertical vibration also applies to the rotation of particles with chiral structure \cite{tsai2005chiral,li2013asymmetric,kummel2013circular,won2019demand}. Other examples of simple self-propelled elements are polar disks \cite{deseigne2010collective}, bolt-like particles \cite{takatori2018cooperative}, walking \cite{patterson2017clogging,deblais2018boundaries,barois2020sorting} and spinning \cite{scholz2018rotating} robots, or asymmetric grains \cite{mohammadi2020dynamics}. The vibration mechanism can also be embedded inside the granular particles as it is for the so-called bristle-bots\cite{ioi1999mobile,giomi2013swarming,koumakis2016mechanism}. In the context of this work, we can also mention some interesting drifting mechanisms with vibrated fluids such as shaken droplets with non-vertical excitation\cite{noblin2009ratchetlike,costalonga2020directional}, droplets on tilted supports\cite{brunet2007vibration,guo2020surface} and walking droplets\cite{couder2005walking,tadrist2018faraday}. 

The above mechanisms for granular systems use elements in a bouncing regime, that mostly have a single mode of locomotion, either linear motion or/and steady rotation, and their structure is approximately rigid or, if not, no mechanical resonances are involved. In this work, we identify a new locomotion mechanism that combines stick-slip transition and mechanical resonance of a soft structure to obtain a controllable motile object with individually addressable forward and backward locomotion states.  
\begin{figure}[h!]
    \centering
    \includegraphics[width=.65\textwidth]{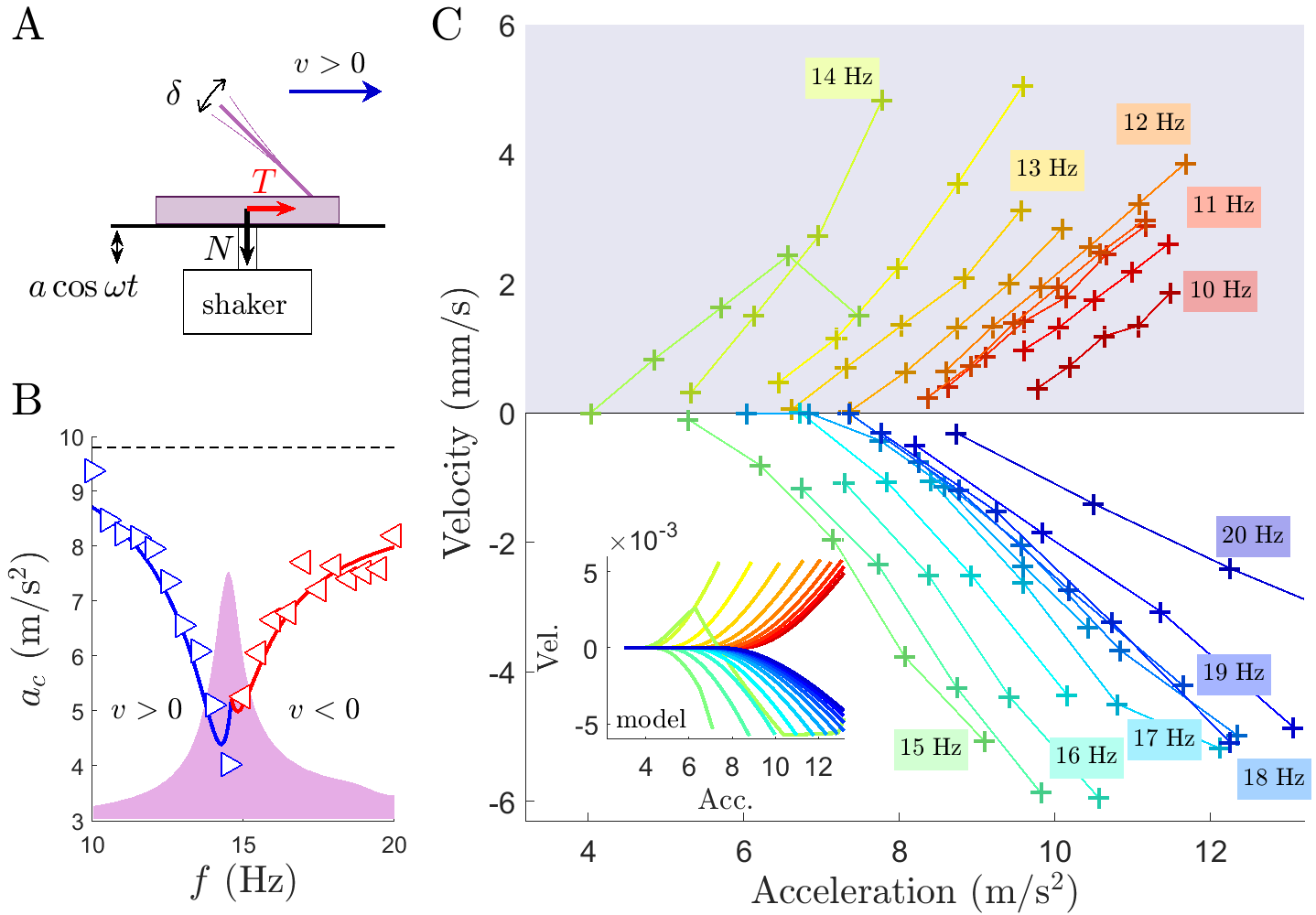}
    \caption{A) Schematics of the locomotion principle. The horizontal black line represents a rigid table sets in vertical oscillatory motion by a shaker at an acceleration amplitude $a$ and frequency $\omega/2\pi$. The soft structure is in frictional contact with the table and it is composed of a rigid base and a tilted soft plate. $N$ is the normal contact force and $T$ is the tangential contact force. $\delta$ is the amplitude of the cantilever first mode. B) Acceleration threshold for steady horizontal motion as a function of the forcing frequency. The solid line is the model in equation (\ref{eq:acc_thresh}) with $\omega_0/2\pi=14.44$~Hz, $Q=17.5$, $g=9.8$ m.s$^{-2}$, $\alpha = 64^\circ$, $M=0.08$ and $c_f=0.25$. Blue: positive velocity, red negative velocity. The resonance response of the soft plate first mode is represented in the background.  C)  
    Velocity as a function of the shaker acceleration $a$. Each set of connected points is for a constant driving frequency. For frequencies from 10 Hz up to 14.5 Hz, the structure moves forward (positive velocities) and for higher frequencies, it moves backward (negative velocities). The inserted plot is the result from the model presented in figure \ref{fig:sol_accel_vs_freq}.}
    \label{fig:velocity_vs_accel}
\end{figure}

Figure \ref{fig:velocity_vs_accel}A represents the essential ingredients for our locomotion mechanism with a structure composed of a rigid base represented by a rectangular block and a tilted flexible plate. The whole structure is placed on a vertically-oscillating rigid table harmonically driven by a shaker at a frequency $\omega/2 \pi$ and an acceleration amplitude $a$. The structure is in solid contact with the table and it can either stick or slide horizontally if the ratio of the tangential forces $T$ to the normal forces $N$ exceed the static friction coefficient. 

Figure \ref{fig:velocity_vs_accel}B represents the threshold acceleration $a_c$ to set the structure in horizontal motion as a function of the frequency. The acceleration threshold is complementary to a resonance-like curve with an acceleration threshold decreasing as the driving frequency approaches the first mode resonance for the structure, which in this example is 14.44 Hz. The sign of the velocity depends on the driving frequency relatively to the  structure internal resonance: for frequencies below the resonance frequency (i.e. between 10 Hz and 14.5Hz), the structure moves forward while above resonance (i.e. between 15 Hz and 20 Hz), the structure moves backward. The threshold acceleration around the resonance is well below the gravity acceleration $g$, which indicates that the structure is in horizontal sliding motion without any jumps. 

Figure \ref{fig:velocity_vs_accel}C represents the structure velocity as a function of the shaker acceleration. Each connected set of data points are for a fixed frequency. For most of the frequencies, the velocity is linear with the acceleration and consistently with figure \ref{fig:velocity_vs_accel}B, the velocity is positive below the resonance and negative above the resonance. For the data points at $f=14.5$ Hz close to the resonance, the threshold acceleration is the lowest at 4 m.s$^{-2}$ but surprisingly the velocity is non-monotonous with the acceleration. 

To explain the net motion of the structure under vibrations and notably the peculiar non-monotonous regime close to resonance, we have to consider the coupling between solid friction and the linear resonance of the structure's first mode. 
The acceleration threshold for the directed motion is related to the threshold condition for sliding in the solid friction regime:
\begin{equation}
\frac{|T|}{|N|} = c_f 		\label{eq:TsN}	
\end{equation}
in which $T$ is the tangential force, $N$ the normal force and $c_f$ the static solid friction coefficient ($T$ and $N$ are represented in figure \ref{fig:velocity_vs_accel}A). Without vibration, there is no tangential force $T=0$ and the normal force is the total weight of the structure of mass $m$, $N=mg$. With the vibration, the net forces $N$ and $T$ have a modulation at the frequency of the shaker with
\begin{eqnarray}
    N &=& mg - ma  \cos \omega t +  m_1\ddot{\delta} \cos \alpha \label{eq:N} \\
    T &=& -m_1 \ddot{\delta} \sin \alpha \label{eq:T}
\end{eqnarray}
in which $m$ is the total mass, $m_1$ is the effective mass of the cantilever first  mode, $a$ is the acceleration amplitude of the shaker and $\alpha$ is the angle between the flapping plate and the horizontal. In equation 2, the term $- a \cos \omega t$ is the modulation of the vertical acceleration in the frame of the vibrating support. $\ddot{\delta}$ is the acceleration of the first mode of the soft plate and it modulates the net forces $N$ and $T$ via the projection with angle $\alpha$. 
The amplitude $\delta$ of the first mode is given by the linear mechanical response $\ddot{\delta} + (\omega_0/Q)\dot{\delta} + {\omega_0}^2 \delta = a   \cos \omega t \cos \alpha$
with $\omega_0/2\pi$ the resonant frequency of the mode and $Q$ the quality factor.
If the sliding condition in equation \ref{eq:TsN} is reached during the time evolution of $N$ and $T$, the disk slides forward or backward, depending on the sign of $T$ when the threshold is reached. 
By combining equations (\ref{eq:TsN}), (\ref{eq:N}) and (\ref{eq:T}), we find two threshold accelerations for the horizontal motion
\begin{equation}
    {a^*}_{\pm} = \frac{g\sqrt{A^2 + B^2}}{A + {\Omega_\pm}^{-2}\left[({\omega_0}^2 - \omega^2)A+ \left(\frac{\omega_0 \omega}{Q}\right)B\right]}\label{eq:acc_thresh}
\end{equation}
with $A = 1 + ({\omega_0}^2 - \omega^2)/{\Omega_\pm}^{2}$ and 
$B = \omega_0\omega/(Q{\Omega_\pm}^2)$ in which $\Omega_\pm$ are two characteristic angular frequencies defined by ${\Omega_\pm}^{-2} =   \omega^2 M\cos \alpha \left(\cos \alpha \pm\sin \alpha/c_f\right)/(({\omega_0}^2 - \omega^2)^2 + \left(\omega_0 \omega/Q\right)^2)$ with $M=m_1/m$ the mass ratio of the resonant mass to the total mass of the structure. The details for the derivation of ${a^*}_{\pm}$ is found in the supplemental material.  The positive solution is for $T=c_fN$ (positive velocity) and the negative solution is for $T=-c_fN$ (negative velocity). In figure \ref{fig:velocity_vs_accel}B, the blue solid line is the solution ${a^*}_{+}$ and the red solid line is the solution ${a^*}_{+}$ without free parameters  (the resonant frequency and the quality factor were independently measured by small amplitude linear response and the friction coefficient was measured by a sliding angle method).
\begin{figure}[h!]
    \centering
    \includegraphics[width=.65\textwidth]{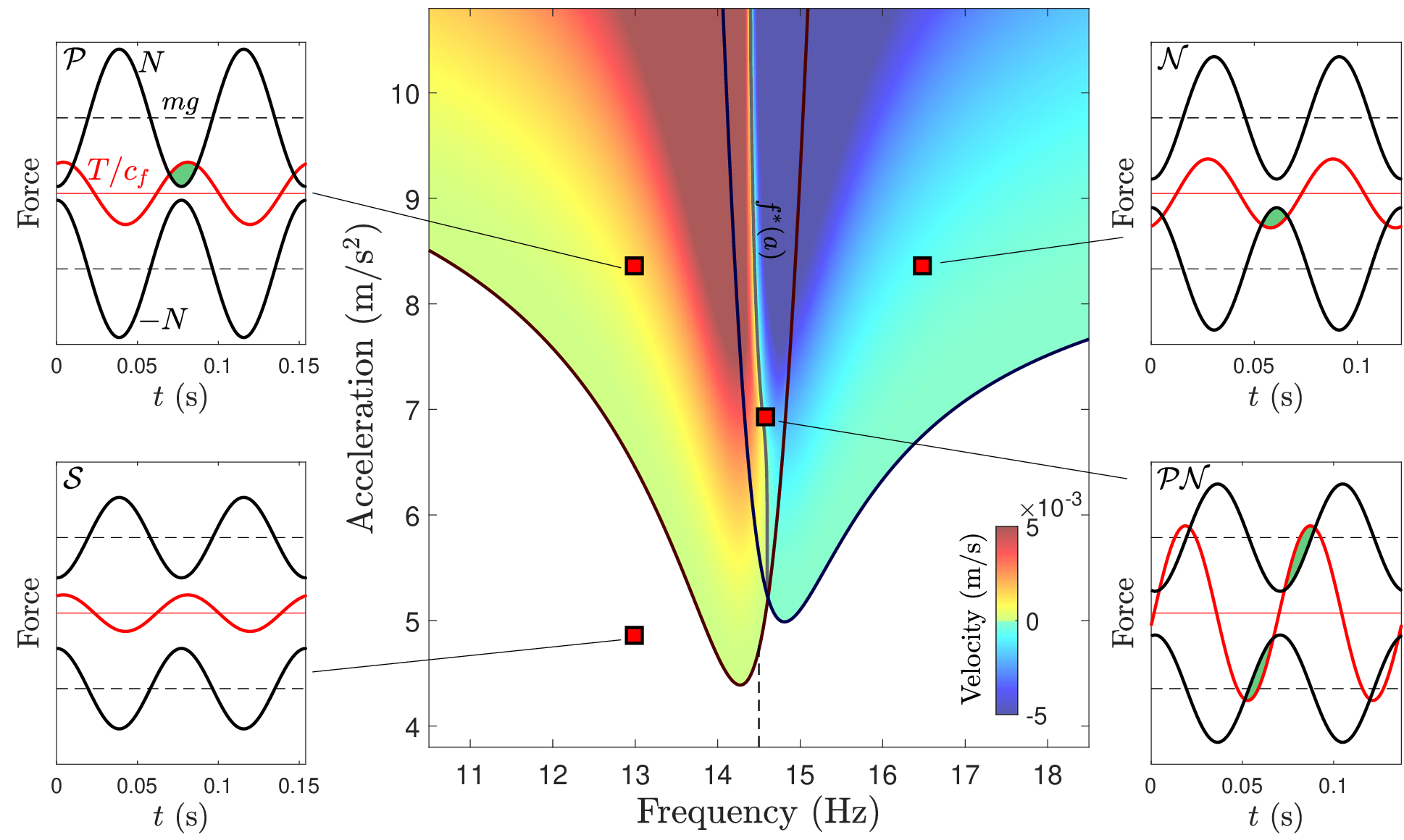}
    \caption{Model for the velocity as a function of frequency and acceleration with 4 regimes (static $\mathcal{S}$, positive $\mathcal{P}$, negative $\mathcal{N}$ and positive-negative $\mathcal{PN}$). The 2 solid lines separating the 4 regimes are the solutions $a^*_\pm$ in equation (\ref{eq:acc_thresh}).  For each of the 4 regimes, we represent the  tangential ($T$) and normal ($N$) forces as a function of time (equations (\ref{eq:N}) and (\ref{eq:T})). The tangential forces are represented with the factor $1/c_f$ with $c_f$ the static friction coefficient so that the intersection between $T/c_f$ and $N$ corresponds to the sliding threshold condition in equation (\ref{eq:TsN}).
In region $\mathcal{P}$ and $\mathcal{N}$, the sliding threshold is reached for positive and negative $T$ respectively.  In the $\mathcal{PN}$ region, the sliding threshold is reached for positive and negative values during a cycle and the frontier $f^*(a)$ between positive and negative net motion is non-vertical.    }
    \label{fig:sol_accel_vs_freq}
\end{figure}

For a driving acceleration larger than the one of the two threshold accelerations ${a^*}_+$ or ${a^*}_-$, the disk has a net displacement during a cycle that we numerically compute by a double time integration of the horizontal net forces using friction laws. The overall sliding distance during an oscillation cycle defines the mean velocity via the period of oscillation $2\pi/\omega$.
Figure~\ref{fig:sol_accel_vs_freq} represents the velocity obtained from this numerical method as a function of the frequency and acceleration of the shaker with a resonant frequency $\omega_0/2\pi=14.44$~Hz, $Q=17.5$, $g=9.8$ m.s$^{-2}$, $\alpha = 64^\circ$, $M=0.08$ and $c_f=0.25$ which corresponds to the experimental values for the structure used in figure \ref{fig:velocity_vs_accel}. 

Figure~\ref{fig:sol_accel_vs_freq} also represents the time evolution of the normal force $N$ and tangential force $T/c_f$, with $c_f$ the static friction coefficient, in the static region $\mathcal{S}$ ($f = 13$ Hz, $a=5$ m.s$^{-2}$), the positive region $\mathcal{P}$ ($f = 13$ Hz and $a=8.5$ m.s$^{-2}$), the negative region $\mathcal{N}$ ($f = 16.5$ Hz and $a=8.5$ m.s$^{-2}$) and the transition region $\mathcal{PN}$ ($f = 14.6$ Hz and $a=7$ m.s$^{-2}$). The time evolution of $N$ and $T$ is given in equations (\ref{eq:N}) and (\ref{eq:T}). In the $\mathcal{S}$  region, $T/c_f$ remains smaller than $N$ which means that the structure is in stiction (no horizontal motion). 
In the positive region $\mathcal{P}$, $T/c_f$ is larger than $N$ during a portion of the cycle which means that the sliding condition (\ref{eq:TsN}) is reached for positive $T$ and the structure slips in the positive direction during each cycle. In the negative region $\mathcal{N}$, the phase shift between $T$ and $N$ is such that  the sliding threshold is reached for $-T/c_f > N$.

In the region $\mathcal{PN}$,  the sliding threshold is reached for positive and negative values and we identify a nonlinear frontier $f^*(a)$ for the velocity cancellation. On this frontier, the structure slips positively and negatively during the cycle but the overall horizontal drift is zero. This frontier results from a nonlinear process: for a frequency such as $f=14.5$ Hz, the horizontal velocity is positive if the acceleration amplitude is $a=6$ m.s$^{-2}$ or negative if the acceleration amplitude is $a=9$ m.s$^{-2}$.  This is a surprising result because the mechanical resonance is in the linear regime which means that the phase shift between $N$ and $T$ is independent of the amplitude $a$. To explain the net velocity sign change close to the resonance, we need to consider the threshold response of the stick-slip transition and how it affects the sliding dynamics of the structure. Close to the resonance, a negative slipping event is immediately followed by a positive slipping event. However, the cumulative sliding distance during one cycle is not a simple combination of the two isolated slipping phases because the initial state of the structure is not the same between the two phases: for the first negative slipping event, the structure is initially in stiction while for the second positive slipping event, the structure has acquired a negative velocity and is initially sliding. We show in more details in the supplemental material how this interaction between the two forcing phases depends on the driving amplitude and can modify the sign of the cumulative horizontal distance close to the resonance, even if the driving frequency is fixed.

\begin{figure}[h!]
    \centering
    \includegraphics[width=.5\textwidth]{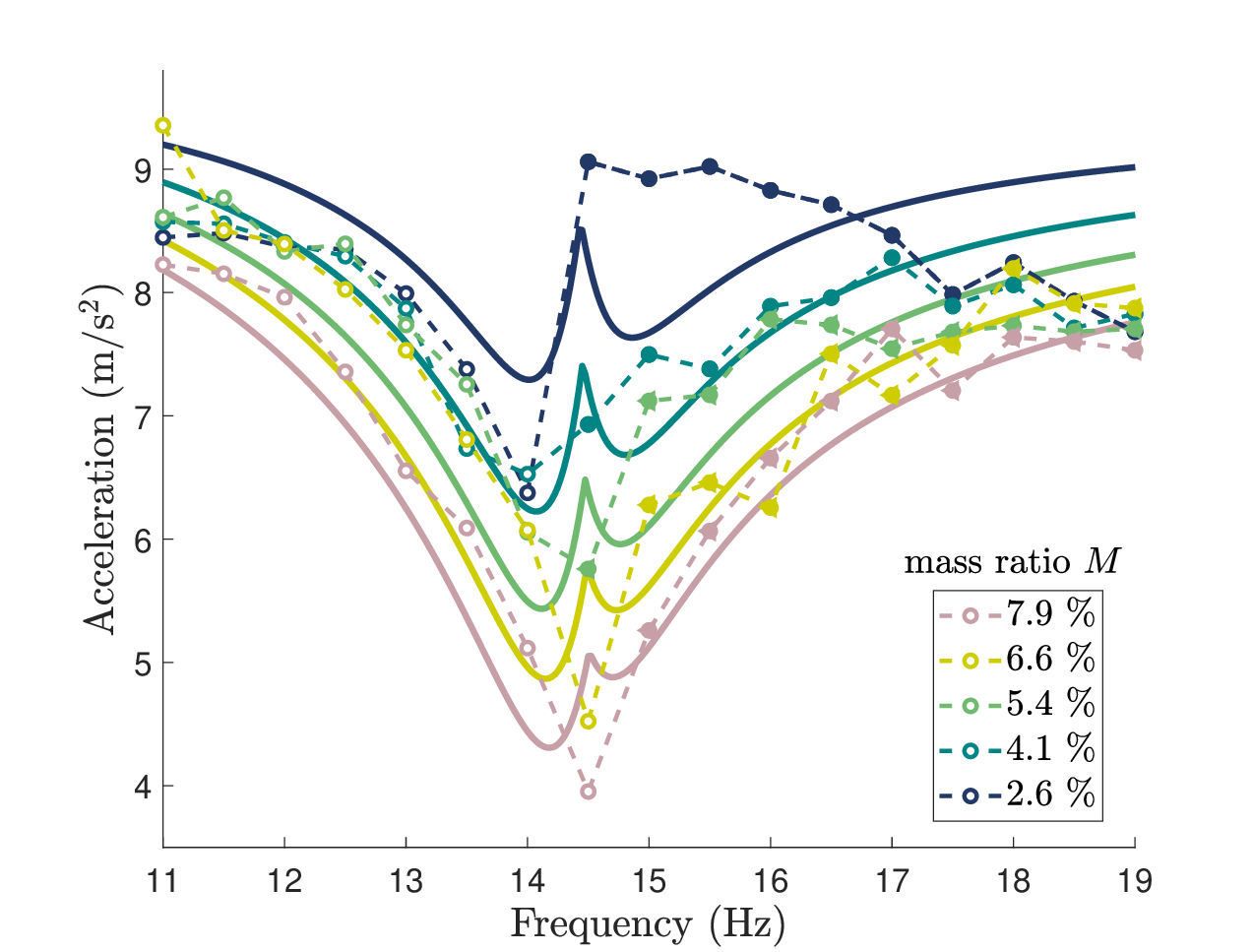}
   \caption{Threshold accelerations of the structure with on-board loads. The structure mass is $m_0=8.2$ g and the added loads are $m_{add} = 0, 1.7, 3.8, 7.7$ and 16.6 g.  
The solid lines are the solutions $a^* = \min \{a^*_\pm\}$  of the model (equation (\ref{eq:acc_thresh})) for the different values of $M=m_1/m$ with $m_1$ the mass of the first mode and $m = m_0 + m_{add}$ the total mass. }
    \label{fig:thresh_vs_mass}
\end{figure}

The control of the horizontal velocity sign is closely related to the fact that the net forces $N$ and $T$ in equations (\ref{eq:N}) and (\ref{eq:T}) oscillate with a different phase. The origin of this phase shift comes from the structure that can be decomposed in a non-resonant mass $m-m_1$, that moves in phase with the oscillating table, and a resonant mass $m_1$, that oscillates with a phase shift because of the harmonic response. In figure \ref{fig:thresh_vs_mass}, we measure the influence of the mass ratio $M= m_1/m$ on the motion threshold. To vary $M$, we add items on the structure base which result in an increase of its non-resonant mass without modifying the resonant mass. We find that the threshold for motion is lowered with increasing mass ratio. For a mass ratio of 6.6\%, the acceleration is typically half the gravity acceleration at the resonance with a quality factor independently measured at $Q=17.5$. This is consistent with the relation $Q m_1 \sim m$ that compares the amplitude of the resonant mass oscillation to the amplitude of the non-resonant mass oscillation. 

Another aspect of the velocity sign control is the behaviour in the limit $M \sim 1$. This limit is when mostly all of the structure mass is resonant. This situation is for example the case for bristle-bots forced by a vibrated table, below the bouncing regime. In this case, the velocity inversion was predicted but experimental results were found inconclusive\cite{cicconofri2016inversion} around the resonant frequency of the structure. Our model shows that, in the limit $M \sim 1$, the region $\mathcal{PN}$ is much wider and the frontier $f^*(a)$ is even more complex than in the case presented in figure \ref{fig:sol_accel_vs_freq} for $M = 0.08$ (see figure 7 in supplemental material).

In figure \ref{fig:deform} we demonstrate that it is possible to combine the resonant oscillation of 2 cantilevers to obtain a fully-controllable structure on a vibrated plate. In figure \ref{fig:deform}A, we represent a structure composed of a rigid disk with two flexible plates (see supplemental material for the structure details). The head plate oscillates in the direction of the head-tail direction and produces a forward or a backward motion depending if the driving frequency is slightly before or after its resonant frequency (14.44 Hz). The second plate is placed at the tail and it oscillates perpendicularly to the head-tail direction. This plate produces a lateral drift force that rotates the structure either clockwise or counterclockwise  depending on the excitation frequency relatively to its resonant frequency (8.5 Hz). The trajectory in figure \ref{fig:deform}A is obtained by vibration of the table following a frequency sequence that successively excites a combination of forward motion (14.2 Hz), right turn (7.2 Hz) and left turn (8.2 Hz) (see movie M1). During the forward motion steps, the structure performs a random walk with a persistence length much larger than its size. By analysing the velocity orientation during the forward motion steps, we find that the angular drift is typically of 0.1 rad for a travelled distance equal to the disk diameter (see supplemental material).
 In figure \ref{fig:deform}B, we show an image sequence of the disk attached to a soft strip (paper strip with the other end perpendicularly fixed to the supporting table at the bottom right corner of the image). By a steady right turn, the structure rolls itself in the strip and ends up in a state in which 3/4 of the disk is covered by the soft strip. The last image of this sequence is the final position reached with the net force from the tail plate balanced by the mechanical stiffness of the soft strip (see Movie M5). In figure \ref{fig:deform}C, we show the bending a soft beam by a combination of forward motions and left turns in order to maintain a perpendicular bending force on the deforming beam. At the end of the sequence, the beam forms a closed loop. In the complete movie sequence (see movie M6), the beam is then released to its initial configuration. 
\begin{figure}[h!]
    \centering
    \includegraphics[width=.75\textwidth]{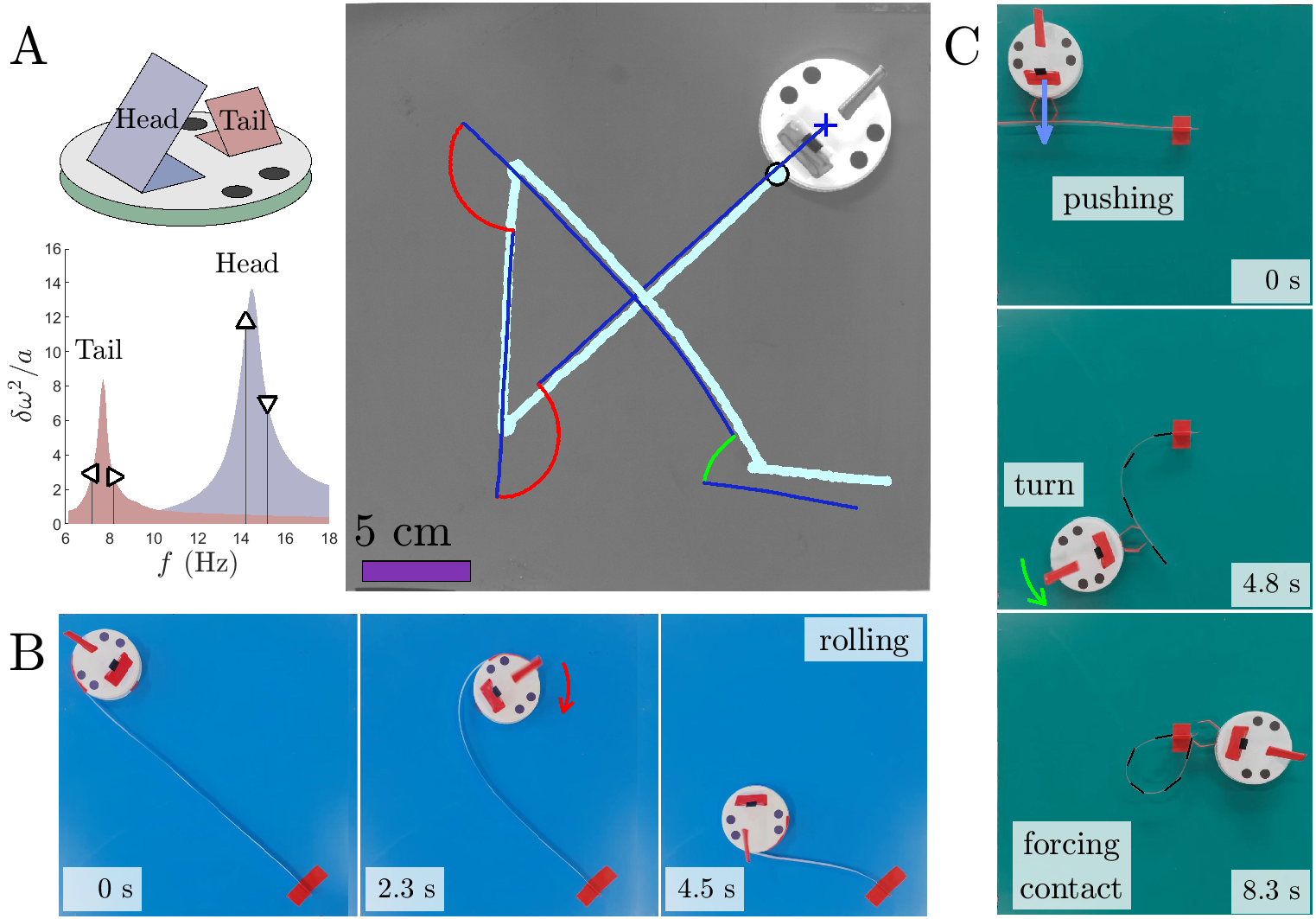}
    \caption{(A) Schematics of a soft structure with two soft plates (head plate for forward and backward motion, tail plate for left and right turns). The trajectory is obtained by sequential excitation of forward motion and left or right turns (see movie M1). The driving frequencies for forward (14.2 Hz), backward (15.2 Hz), left (7.2 Hz) and right turns (8.2 Hz) are indicated on the response function plot with triangular symbols pointing respectively up, down, left and right. (B) Image sequence with the structure attached to a paper strip and rolling itself via a steady right turn (see movie M5). (C) Image sequence showing the bending of a soft beam to a self-contact (see movie M6). }
    \label{fig:deform}
\end{figure}

\begin{figure}[h!]
    \centering
    \includegraphics[width=.8\textwidth]{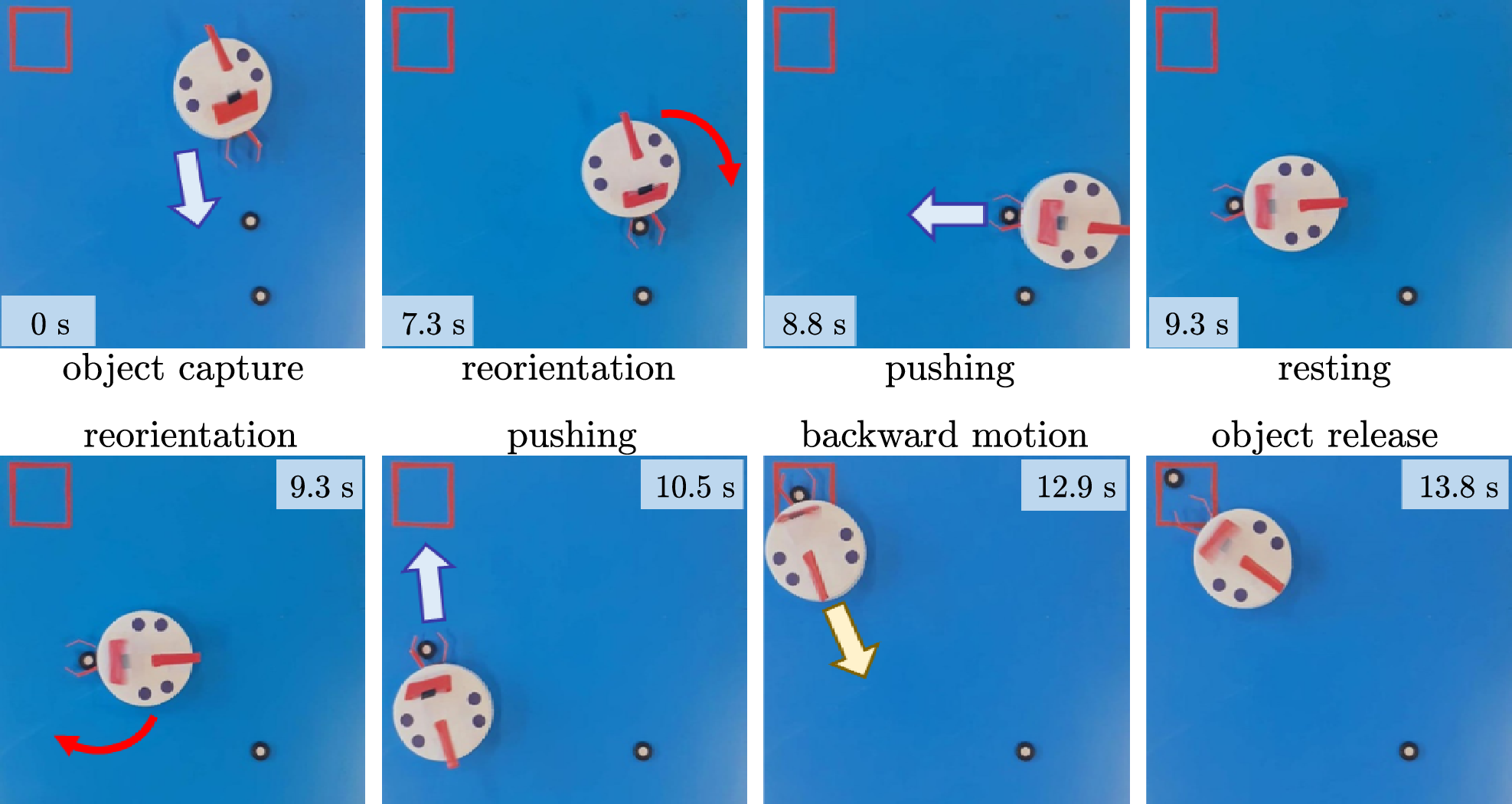}
    \caption{Capture and delivery of 2 small cylinders (diameter 1.2 cm) into a target area. The movie M7 shows the full sequence.}
    \label{fig:delivery}
\end{figure}

In the sequence of figure \ref{fig:delivery}, we demonstrate how the structure can move small objects from some location on the supporting plate towards a target area identified by a red square (see movie M7 for the complete sequence). 

In this work, we identified a locomotion mechanism that uses simple and universal physical ingredients: solid friction and mechanical resonance. The coupling between the resonant structure and the stick-slip dynamics of a solid body subjected to vibrations gives rise to rectified motion in both the forward and backward directions. The use of multiple flaps further allows to give rise to new degrees of freedom leading to moving and rotating structures that can be controlably steered. This easily-applicable mechanism offers new possibilities for the investigation of active matter beyond directional motion or for the development of soft and minimal robotic structures\cite{shepherd2011multigait,rubenstein2014kilobot,marvi2014sidewinding,rus2015design,aguilar2016review,laschi2016soft,mintchev2016adaptive,hawkes2017soft,gu2018soft,hu2018small,mahon2019soft,wu2019insect,duduta2020tunable,liang2021electrostatic,ben2023morphological,chong2023multilegged,boudet2021collections}.

\end{document}